\documentclass[12pt]{article}
\usepackage{geometry} 
\usepackage{tocbibind}
\usepackage[titletoc,title]{appendix}
\usepackage{fancyhdr}
\usepackage[usenames,dvipsnames]{color}
\usepackage{parskip}
\usepackage{graphicx}
\usepackage{hyperref}
\usepackage[affil-it]{authblk}
\usepackage{mathtools}
\usepackage{kbordermatrix}
\usepackage{amsmath}
\usepackage{bbm}
\usepackage{tensor}
\usepackage{braket}
\usepackage{subcaption}
\usepackage{caption}
\usepackage{calrsfs}
\usepackage{xspace}

\newcommand{\Poincare}{Poincar\'e\xspace}
\newcommand*\dd{\mathop{}\!\mathrm{d}}

\newcommand{\beq}{\begin{equation}}
\newcommand{\eeq}{\end{equation}}

\newcommand{\abs}[1]{\left| #1 \right|}
\DeclarePairedDelimiter{\evdel}{\langle}{\rangle}
\newcommand{\ev}{\evdel}

\fancypagestyle{firststyle}
{
   \fancyhead[RO]{SLAC-PUB-16904}
   \fancyfoot[C]{\footnotesize  \thepage\ }
}

\title{\bf\Large{Angular Momentum Sum Rules in the Front Form}}

\begin{document}
\title{\bf\Large{Angular Momentum Conservation Law in Light-Front Quantum Field Theory}}
\author[1]{Kelly Yu-Ju Chiu\thanks{yujuchiu@stanford.edu}}
\author[1]{Stanley J. Brodsky\thanks{sjbth@slac.stanford.edu}} 
\affil[1]{\normalsize{SLAC National Accelerator Laboratory,

 Stanford University, Stanford, California 94039, USA}} 
\setlength{\affilsep}{1em}


\maketitle
\abstract{We prove the Lorentz invariance of the angular momentum conservation law and the helicity sum rule for relativistic composite systems in the light-front formulation.  We explicitly show that $j^3$, the $z$-component of the angular momentum remains unchanged under  Lorentz transformations generated by the light-front kinematical boost operators. The invariance of $j^3$ under Lorentz transformations is a feature unique to the front form. 
Applying the Lorentz invariance of the angular quantum number in the front form, we obtain a selection rule for the orbital angular momentum which can be used to eliminate certain interaction vertices in QED and QCD. We also generalize the selection rule to any renormalizable theory and show that there exists an upper bound on the change of orbital angular momentum in scattering
processes at any fixed order in perturbation theory.
}
\thispagestyle{firststyle} 
\cleardoublepage

\section{Introduction}
Understanding the angular momentum decomposition and helicity sum rule for nucleons  is of great interest in hadron physics. One of the difficulties  in studying this problem is the non-uniqueness of the definition of relativistic spin \cite{Bakker:2004ib} \cite{Carlson:2003je} \cite{Polyzou:2012ut}. To avoid this issue, nucleons are usually studied in a frame in which the nucleons move along the $z$-direction such that helicity coincides with the $z$-projection of spin, since it is generally believed that the helicity and spin are not the same in an arbitrary frame  \cite{Leader:2013jra}. Since constituents in a bound state can move in different directions, it is understood that for the Wick helicity spin states, there is no conservation law of helicity.

Nonetheless, as we shall show in this paper, the
$z$-component of the relativistic spin of a particle or bound state in the front form \cite{RevModPhys.21.392} is Lorentz invariant and always equal to its helicity. Furthermore, we will prove that  for any composite system,  helicity is conserved in any Lorentz frame. This is related to the fact that Lorentz transformations in the front form are generated by \textit{kinematical} operators which leave the $x^+=0$ plane invariant, whereas boosts in the instant form are \textit{dynamical} and the $x^0=0$ plane is changed under Lorentz transformations \cite{Kogut:1969xa} \cite{Soper:1972xc} \cite{Leutwyler:1977vy}. The invariance of spin in the front form provides selection rules for orbital angular momentum in interaction vertices and scattering  processes in renormalizable theories. Examples of the selection rules have been observed in \cite{Cruz-Santiago:2015dla} \cite{Brodsky:2000ii} \cite{Dixon:1996wi}.

The paper is organized as follows: In Section 2, we briefly remind the readers why spin in relativistic theories is nontrivial, and which are the different definitions of relativistic spin states generally used in the literature.  In Section 3, we compare the dependence of the expectation value of spin operators on different choices of spin states, and show that the light-front spin choice is unique:  the spin expectation value along the $z$-direction is always conserved under Lorentz transformations.  We then give a general proof for the Lorentz invariance of angular momentum along the $z$-direction in the front form for both elementary and composite particles.  In Section 4, we present a selection rule for the angular momentum in QED and QCD vertices by applying the light-front angular momentum conservation law. We also give an upper bound on the change of orbital angular momentum in scattering processes for renormalizable theories at any fixed order in perturbation theory.   Conclusions are summarized in Section 5.

For completeness and clarity, we also include in  Appendix (A) light-front conventions and a glossary of notations which we use, (B) derivation of light-front spin representations, (C) relations between light-front spin operators, the covariant spin vector and the Pauli-Lubanski pseudovector.

\section{Spin of Relativistic Particles}
In $d=3+1$ dimensions, the \Poincare group has two Casimir operators, $P^2= m^2$ and $ W^2$, where $ W^\mu= -\frac{1}{2} \varepsilon^{\mu\nu\alpha\beta} { P}_\nu { M}_{\alpha\beta} $ is the Pauli-Lubanski pseudovector. For a fixed momentum $ p^\mu$,
$W^\mu$ is the generator of the little group, the maximal subgroup of the Lorentz group which leaves $ p^\mu$ invariant. According to Wigner's theorem, elementary particles classified with $m^2$ and $W^2$ transform  in  unitary irreducible representations of the symmetry group.

In the following, we will construct the spin representations for both massive and massless elementary particles, respectively. In both cases, we shall start with a standard reference frame in which the spin is unambiguously defined, and then apply Lorentz transformations to obtain the spin in any arbitrary Lorentz frame. Since the Lorentz transformation between  two frames is not unique, we then discuss the spin states defined by  different choices of Lorentz transformations.

\subsection{Massive elementary particles }
For massive elementary particles, the intuitive choice for the standard reference frame is the rest frame, in which the momentum is ${\mathring{p}}^\mu \equiv\kbordermatrix{&0&1&2&3\\ &m&0&0&0}$. The Pauli-Lubanski pseudovector in this frame is
\beq
 W^\mu= m \begin{bmatrix}
0\\ 
J^1\\
J^2\\
J^3
\end{bmatrix}= m \begin{bmatrix}
0\\ 
{S}^i
\end{bmatrix},\;\; i=1,2,3,  \label{1}
\eeq
where $J^i= L^i+ S^i=\frac{1}{2}\epsilon^{ijk}M^{jk}$ are the total rotation generators in $3$ dimensions. ${S}^i$ are the spin generators, and the orbital generators $L^i$ do not contribute when particles are at rest\footnote{Throughout this paper, we will reserve uppercase letters for operators, and use lowercase letters to denote the value of the operator acting on some states. For example, the momentum operator on a momentum eigenstate is denoted by $P^\mu\ket{p}= p^\mu \ket{p}$.  A full gloassary of symbols is given in Appendix A.1.}.  The Casimir  $W^2= -m^2 {(S^i)}^2=-m^2 s (s+1)$ is \Poincare invariant, and $s$ is defined as the spin representation of a particle in a relativistic theory \cite{murayama} \cite{Schwartz:2013pla}. 
 
In the rest frame of a spin-$s$ massive particle, the spin is uniquely labeled by  $s^3$, the $(2s+1)$ eigenvalues  along the $z$-direction, which we will use interchangeably with helicity $\lambda$: 
\beq
S^3 \ket{\mathring{p}; \lambda=s^3}= \lambda  \ket{\mathring{p}; \lambda} \text{\;\; for } \lambda= -s, -s+1,...,0,...,s-1, s. \label{so3}
\eeq

Although the spin is well-defined in the particle's rest frame, the definition of spin for a particle in motion  is convention dependent, since in fact a Lorentz transformation from the rest frame $\ket{\mathring{p}}$ to a state $\ket{p}$ with momentum is not unique. Generally speaking,  a particle with spin in the $z$-direction and $s^3=\lambda$ in its rest frame  is not guaranteed to have the spin aligned in the $z$-direction when it is moving. Therefore, even though the helicity $\lambda$ is a Lorentz invariant label of particles, it should not be identified with the $z$-component of spin for particles in motion.  Nonetheless, as we will see in Section 2, there is a particular choice of Lorentz transformation under which, spin-projection along the $z$-direction is invariant, and helicity is equal to the $z$-component of spin in all Lorentz frames.

There are three popular choices of Lorentz transformations in the literature which give rise to different definitions of relativistic spin states \cite{Carlson:2003je} \cite{Polyzou:2012ut} \cite{Leader:2001gr}\cite{Li:2015hew}: \\

\begin{enumerate}
\item \underline{Canonical spin} 
Starting with the rest frame of a massive particle in which the spin is projected along the $z$-direction, the {canonical} spin states are obtained by first performing a rotation from the direction of $\mathbf p$ to the $z$-axis,  followed by a boost along the $z$-direction to get the desired $ \abs{\mathbf p}$, and finally a rotation from the $z$-axis back to the $3$-momentum direction $\mathbf p$:
\begin{align}
\ket{p; \lambda}_c & \equiv \Lambda_c(\mathring{p} \rightarrow p)  \ket{\mathring{p}; s^3= \lambda}\\[5pt]
& =R(\hat z \rightarrow \hat{\mathbf p})\; B(\mathring p \rightarrow p^3= \abs{\mathbf p})\; R^{-1}(\hat z \rightarrow \hat{\mathbf p})\ket{\mathring{p}; s^3= \lambda} \label{lac},
\end{align}
where
\beq
R(\hat z \rightarrow \hat{\mathbf p})=e^{-i M^{12}\phi} e^{-i M^{31}\theta},\;\;  \phi= tan^{-1}\frac{p^1}{p^2},\; \theta= tan^{-1}\frac{\sqrt{{(p^1)}^2 + {(p^2)}^2}}{p^3} \label{R}
\eeq
and
\beq
B(\mathring p \rightarrow p^3= \abs{\mathbf p})=  e^{-i M^{03}\rho},\;\;  \rho= tanh^{-1}\frac{\abs{\mathbf p}}{p^0} \label{B}.
\eeq
Note that the action of $\Lambda_c$ is equivalent to a \textit{rotationless pure boost} along  the direction of the $3$-momentum  $\mathbf{p}$. \\


The $4$-vector representation of $\Lambda_c$ is given by
\begin{align}
\tensor{\left(\Lambda_c\right)}{^\mu_\nu}(\mathring{p} \rightarrow p)  
  &= \kbordermatrix{
    & 0 & i \\
    0 &  \frac{p^0}{m}       &    \frac{p^i}{m} \\
    i& \frac{p^i}{m}        &  \delta^{ij}+ \frac{p^i p^j}{m(p^0+m)}}\\[5pt]
&=\kbordermatrix{
    & 0 & 1& 2 & 3 \\
          0 & \frac{p^0}{m}       & \frac{p^1}{m}& \frac{p^2}{m} & \frac{p^3}{m}  \\\\
          1 &  \frac{p^1}{m}       & 1 + \frac{p^1 p^1}{m(p^0+m)}& \frac{p^1 p^2}{m(p^0+m)} & \frac{p^1 p^3}{m(p^0+m)} \\\\
          2 & \frac{p^2}{m}       & \frac{p^2p^1}{m(p^0+m)}& 1+ \frac{p^2 p^2}{m(p^0+m)} & \frac{p^2 p^3}{m(p^0+m)} \\\\
          3 & \frac{p^3}{m}       & \frac{p^3p^1}{m(p^0+m)}&  \frac{p^3 p^2}{m(p^0+m)} & 1+ \frac{p^3 p^3}{m(p^0+m)} 
  } ,
\label{vecc}
\end{align}
with $p^0= \sqrt{{\abs{\mathbf p}}^2 + m^2}$ for particles which are on-shell. 

Using Eq.(\ref{vecc}), we see that for a particle polarized along the $z$-direction in its rest frame with the  covariant\footnote{A detailed discussion on the covariant spin vector and its relation to the Pauli-Lubanski pseudovector can be found in Appendix C.1.} spin $4$-vector  $s^\mu= \kbordermatrix{&0&1&2&3\\ &0&0&0&m}$, after performing the canonical choice of Lorentz transformation, the spin in general will not be aligned with  the  $3$-momentum $\mathbf p$. 

In the low-energy limit $(\abs{\mathbf p} \ll m) $,  the canonical spin defined by $\Lambda_c$ is the natural choice since $\Lambda_c$ is smoothly connected to the identity, and the spin is unchanged under Galilean boosts, as expected in non-relativistic physics. \\

\item \underline{Wick helicity spin} Helicity states are defined such that the spin of the moving particle is parallel or anti-parallel to the direction of the\\ $3$-momentum $\mathbf{p}$. 

Starting with a massive particle in the rest frame, helicity states are obtained by  boosting along the $z$-direction to obtain the desired $ {\mathbf p}$, followed by a rotation from the $z$-axis to the direction of $\abs{\mathbf p}$:
\begin{align}
\ket{p; \lambda}_h &\equiv \Lambda_h(\mathring{p} \rightarrow p)  \ket{\mathring{p}; s^3= \lambda}\\[5pt]
&=R(\hat z \rightarrow \hat{\mathbf p})\; B(\mathring p \rightarrow p^3= \abs{\mathbf p})\ket{\mathring{p}; s^3= \lambda}, \label{lah}
\end{align}
where $R(\hat z \rightarrow \hat{\mathbf p})$ and  $B(\mathring p \rightarrow p^3= \abs{\mathbf p})$ are defined in Eq.(\ref{R}) and (\ref{B}). Unlike the canonical choice,   no  rotation is performed in the rest frame in the helicity choice before boosting in the $z$-direction. This ensures that the rest-frame spin vector which is pointed along the $z$-direction 
 will be aligned with the $3$-momentum $\mathbf p$ after the helicity boost.

Helicity boost is related to the canonical boost\footnote{For elementary particles, any two choices of Lorentz transformations are related to each other by a pure rotation, known as the generalized Melosh rotation. The reason is the following. Assume there are two boosts, $\Lambda_A$ and $\Lambda_B$, both of which transform a particle at rest to a state with momentum $p$, and $
{p}^\mu= \tensor{\left(\Lambda_A\right)}{^\mu_\nu} \mathring{p}^\nu= \tensor{\left(\Lambda_B\right)}{^\mu_\nu} \mathring{p}^\nu
$. It then follows that \\
$\mathring{p}^\mu= \tensor{\left(\Lambda^{-1}_A\right)}{^\mu_\nu} p^\nu= \tensor{\left(\Lambda^{-1}_A\right)}{^\mu_\alpha}\tensor{\left(\Lambda_B\right)}{^\alpha_\nu} \mathring{p}^\nu.
$
Since $\mathring{p}$ has vanishing space component,  in order for the equality to hold, $\Lambda^{-1}_A \Lambda_B$ can only be a pure spatial rotation. }  by \\ $\Lambda_h(\mathring{p} \rightarrow p)= \Lambda_c(\mathring{p} \rightarrow p) R(\hat z \rightarrow \hat{\mathbf p})$.\\

The $4$-vector representation of $\Lambda_h$ is given by
\beq 
\tensor{\left(\Lambda_h\right)}{^\mu_\nu}(\mathring{p} \rightarrow p) = \kbordermatrix{
 & 0 & 1& 2 & 3 \\
    0& \frac{p^0}{m}       & 0	 & 	0  & \frac{\abs{\mathbf p}}{m}  \\\\
    1& \frac{p^1}{m}       &  \frac{p^1 p^3}{\abs{\mathbf p}\abs{ p^\perp}}& \frac{-p^2}{\abs{ p^\perp}} & \frac{p^0 p^1}{m\abs{\mathbf p}}\\\\
     2& \frac{p^2}{m}       & \frac{p^2 p^3}{\abs{\mathbf p}\abs{ p^\perp}}&  \frac{p^1}{ \abs{ p^\perp}} & \frac{p^0 p^2}{m\abs{\mathbf p}}\\\\
       3& \frac{p^3}{m}       & \frac{-\abs{p^\perp}}{\abs{\mathbf p}}& 0 & \frac{p^0 p^3}{m\abs{\mathbf p}} 
       }, \label{vech}
\eeq
with ${\abs{p^\perp}}^2= {(p^{\perp})}^2 ={(p^1)}^2 + {(p^2)}^2$.

Using Eq.(\ref{vech}), it is obvious that for a particle polarized along the $z$-direction in its rest frame with the covariant spin   spin $4$-vector \\ $s^\mu= \kbordermatrix{&0&1&2&3\\ &0&0&0&m}$, after the helicity boost, the spin will be aligned with the $3$-momentum $\mathbf p$.\\

\item \underline{Light-front spin} Light-front states are defined using the light-front \textit{kinematical} boost generators, $M^{+\perp}$ and $M^{+-}$, which leaves the $x^+=0$ plane invariant; this is in  contrast to the canonical or Wick helicity boost in the instant form, where boost generators are \textit{dynamical} and  the $x^0=0$ plane is changed under Lorentz transformations.  Note that it is constructed such that  the direction of spin in the particle's rest frame coincides with the light-front direction\footnote{Light-front conventions which we use are listed in Appendix A.2}  $\hat z$, and as we will see, this choice makes the $z$-component of spin special in Lorentz transformations.

Light-front states are obtained from the rest frame of a massive particle by first boosting in the $z$-direction to obtain the desired $ p^+$, followed by a light-front transverse boost from the $z$-axis to obtain the desired transverse momentum $p^\perp$:
\begin{align}
\ket{p; \lambda}_L &\equiv \Lambda_L(\mathring{p} \rightarrow p)  \ket{\mathring{p}; s^3= \lambda} \label{ti_l}\\[5pt]
&=e^{-i M^{+ \perp}\theta^\perp} e^{-i \frac{M^{+-}\omega}{2}}\ket{\mathring{p}; s^3= \lambda}  \label{lal}
\end{align}
where
\begin{align}
&\theta^\perp=\frac{p^\perp}{p^+}, \;\;\perp=1,2 \label{13}
\\   &e^\omega=\frac{m}{p^+} . \label{14}
\end{align}

The $4$-vector representation of $\Lambda_L$ is given by
\beq 
\tensor{\left(\Lambda_L\right)}{^\mu_\nu}(\mathring{p} \rightarrow p) = \kbordermatrix{
    & + & - & 1 & 2 \\
     + & \frac{p^+}{m}       & 0	 & 	0  & 0 \\\\
   -& \frac{{|p^\perp|}^2}{m p^+}& \frac{m}{ p^+} & \frac{2 p^1}{p^+} & \frac{2 p^2}{p^+}\\\\
     1& \frac{p^1}{m}       & 0 & 1 & 0\\\\
      2&  \frac{p^2}{m}       & 0 & 0 & 1
}. \label{vecl}
\eeq
In the light-front boost, the parameters $(\theta^\perp, \omega)$  have simple connections\footnote{This simplification occurs because kinematical generators of the \Poincare group on the light-front are isomorphic to the symmetry operators of non-relativistic quantum mechanics in $d=2+1$ dimensions\cite{Kogut:1969xa}.} to the momentum $p$, in contrast to the canonical or helicity boosts, where parameters $(\theta, \phi, \rho)$ are non-linear functions of the momentum as in Eq. (\ref{R}) and (\ref{B}). 

We see that the light-front states are similar to  the Wick helicity states in the sense that the spin of the moving particle will be parallel or anti-parallel to the light-front $3$-momentum $(p^+, p^\perp)$.\\
\end{enumerate}

\begin{figure}[h]
\centering
\includegraphics[width=0.28\textwidth]{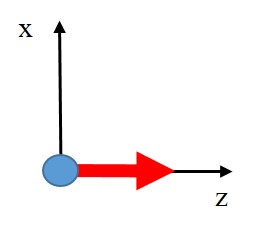}
\caption{Spin in the rest frame is aligned with the $z$-direction}
\label{rest}
\end{figure}

\begin{figure}[h]
\centering
\begin{subfigure}[t]{.28\textwidth}
\centering
\captionsetup{justification=centering}
  \includegraphics[width=\textwidth]{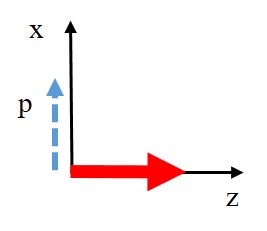}
  \caption{Canonical spin\\ $s^\mu_c(p)=(0 ,0,0,m)$}
  \label{spins_a}
\end{subfigure}%
\begin{subfigure}[t]{.3\textwidth}
\centering
\captionsetup{justification=centering}
  \includegraphics[width=\textwidth]{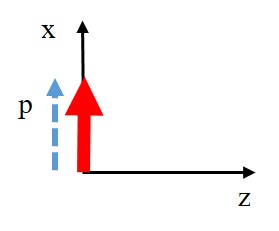}
  \caption{Helicity spin \\$s^\mu_h(p)=(p ,E,0,0)$}
    \label{spins_b}
\end{subfigure}%
\begin{subfigure}[t]{.28\textwidth}
\centering
\captionsetup{justification=centering}
  \includegraphics[width=\textwidth]{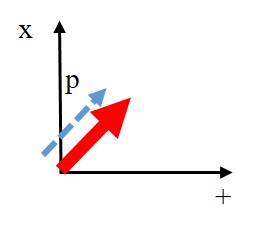}
  \caption{Light-front spin \\$s^\mu_L(p)=(\frac{p^2}{E} ,p,0,\frac{m^2}{E})$}
    \label{spins_c}
\end{subfigure}%
\caption{ Different definitions of  spin for a massive particle moving with momentum $p^\mu=(E,p,0,0)$; only the spatial components of the covariant spin $4$-vector are represented in the figures. The covariant spin vector $s^\mu(p)$ is written in the $(0,1,2,3)$ coordinates. Note that in Fig.\ref{spins_c} the horizontal axis is the $+$ direction. }
\label{spins}
\end{figure}

We give a graphical illustration below on how a covariant spin $4$-vector appears under the different choices of Lorentz transformations. In the example, we consider a massive  particle traveling along the $x-$direction with  $p^\mu= \kbordermatrix{&0&1&2&3\\ &E&p&0&0}$. The spin  is originally polarized along the $z$-direction in the rest frame with $s^\mu= \kbordermatrix{&0&1&2&3\\ &0&0&0&m}$, as shown in Fig.\ref{rest}. The spin states corresponding to the different choices of Lorentz transformations are illustrated in Fig.\ref{spins}.

\subsection{Massless elementary particles}
Unlike  the massive case, for massless particles, spin is not directly defined from the eigenvalues of $W^2$. This is because the little group of the \Poincare  group for massless particles is the non-compact isometry group of the $2-$dimensional Euclidean space  $ISO(2)$, which does not admit finite-dimensional unitary representations. However, particles are observed to have discrete spin quantum numbers in addition to momentum $p$. Thus, all non-compact generators of $ISO(2)$ are neglected, and the remaining generators form a compact $SO(2)$ group.

For each spin-$s$ irreducible representation of the $SO(2)$ group, there are only two linearly independent polarization states with eigenvalues $s$ and $-s$, respectively. States corresponding to the two eigenvalues are referred to as the $``+"$ and $``-"$ helicity states of massless particles, because the $SO(2)$ generator points along the direction of $\mathbf p$.
 
Since spin states are labeled by the $S^3$ operator in the massive case, to be consistent,  one defines spin in a frame in which the massless particle moves along the $z$-direction with momentum ${\bar p}^\mu\equiv \kbordermatrix{&0&1&2&3\\ &\bar p&0&0&\bar p}$ so that the $SO(2)$ group in this frame is generated by $S^3$. Spin states for massless particles are then labeled by 
\beq
S^3 \ket{\bar{p}; \lambda=s^3}= \lambda  \ket{\bar{p}; \lambda} \text{\;\; for } \lambda= \pm s. \label{so2}
\eeq

Now that we have defined the spin for massless particles moving along the $z$-direction, we can construct spin states for massless particles moving with arbitrary momentum $p$ as we did in the massive case. However the \textit{canonical} spin definition is not suitable for massless particles because it requires a rest frame from which a pure boost is performed and there is no rest frame for massless particles. 

In the following, we will discuss the remaining two choices of Lorentz transformations for massless particles.\\
\begin{enumerate}
\item \underline{Wick helicity spin}  Helicity spin states for massless particles  are obtained by first boosting a state with momentum $\bar p$ in the $z$-direction to obtain the desired $ \abs{\mathbf p}$, and then rotating from the $z$-axis to the direction of ${\mathbf p}$ to have the desired the transverse momentum:
\begin{align}
\ket{p; \lambda}_h & \equiv \Lambda_h(\bar{p} \rightarrow p)  \ket{\bar{p}; s^3= \lambda}\\[5pt]
& =R(\hat z \rightarrow \hat{\mathbf p})\; B(\bar p \rightarrow p^3= \abs{\mathbf p})\; \ket{\bar{p}; s^3= \lambda}, 
\end{align}
where
\begin{align}
&R(\hat z \rightarrow \hat{\mathbf p})=e^{-i M^{12}\phi} e^{-i M^{31}\theta},\;\;  \phi= tan^{-1}\frac{p^1}{p^2},\; \theta= tan^{-1}\frac{\abs{p^\perp}}{p^3} 
\end{align}
and
\begin{align}
&B(\bar p \rightarrow p^3= \abs{\mathbf p})=  e^{-i M^{03}\rho},\;\;  e^\rho= \frac{\abs{\mathbf p}}{\bar p}.
\end{align}

The $4$-vector representation of $\Lambda_h$ for massless particles is given by
\beq 
\tensor{\left(\Lambda_h\right)}{^\mu_\nu}(\bar{p} \rightarrow p) = \kbordermatrix{
 & 0 & 1& 2 & 3 \\
    0& \frac{{\abs{\mathbf p}}^2 + {(\bar p)}^2}{\abs{\mathbf p}\bar p }       & 0	 & 	0  & \frac{{\abs{\mathbf p}}^2 - {(\bar p)}^2}{\abs{\mathbf p}\bar p }  \\\\
    1& \frac{p^1\left({\abs{\mathbf p}}^2 - {(\bar p)}^2\right) }{{\abs{\mathbf p}}^2\bar p }      &  \frac{p^1 p^3}{\abs{\mathbf p}\abs{ p^\perp}}& \frac{-p^2}{\abs{ p^\perp}} & \frac{p^1\left({\abs{\mathbf p}}^2 + {(\bar p)}^2\right) }{{\abs{\mathbf p}}^2\bar p }     \\\\
     2& \frac{p^2\left({\abs{\mathbf p}}^2 - {(\bar p)}^2\right) }{{\abs{\mathbf p}}^2\bar p }          & \frac{p^2 p^3}{\abs{\mathbf p}\abs{ p^\perp}}&  \frac{p^1}{ \abs{ p^\perp}} & \frac{p^2\left({\abs{\mathbf p}}^2 + {(\bar p)}^2\right) }{{\abs{\mathbf p}}^2\bar p }   \\\\
       3& \frac{p^3\left({\abs{\mathbf p}}^2 - {(\bar p)}^2\right) }{{\abs{\mathbf p}}^2\bar p }          & \frac{-\abs{p^\perp}}{\abs{\mathbf p}}& 0 & \frac{p^3\left({\abs{\mathbf p}}^2 + {(\bar p)}^2\right) }{{\abs{\mathbf p}}^2\bar p }   
       }. \label{vec_h}
\eeq
Note that in general, $\abs{{\mathbf p}}^2 \neq {(\bar p)}^2$ due to the non-unitarity of the boost operation.

The above expression shows that after the helicity boost, the spin vector in the standard reference frame indeed transforms into a vector which points in  the $3$-momentum  direction $\mathbf p$. \\

\item \underline{Light-front spin} Light-front spin states  for massless particles are obtained by boosting $\bar p$ in the $z$-direction to obtain the desired $ p^+$, followed by a light-front transverse boost from the $z$-axis to obtain the desired transverse momentum $p^\perp$. 
\begin{align}
\ket{p; \lambda}_L &\equiv \Lambda_L(\bar{p} \rightarrow p)  \ket{\bar{p}; s^3= \lambda}\\
=&e^{-i M^{+ \perp}\theta^\perp} e^{-i \frac{M^{+-}\omega}{2}}\ket{\bar{p}; s^3= \lambda}
 \end{align}
where
\begin{align}
&\theta^\perp=\frac{p^\perp}{p^+},\;\; \perp=1,2 \label{255} \\   &e^\omega=\frac{2 \bar p}{p^+}.
\label{256}
\end{align}

The  $4$-vector representation of $\Lambda_L$ for massless particles is given by
\beq 
\tensor{\left(\Lambda_L\right)}{^\mu_\nu}(\bar{p} \rightarrow p) = \kbordermatrix{
    & + & - & 1 & 2 \\
     + & \frac{p^+}{2 \bar p}       & 0	 & 	0  & 0 \\\\
   -& \frac{{|p^\perp|}^2}{2 \bar p p^+}& \frac{2 \bar p}{ p^+} & \frac{2 p^1}{p^+} & \frac{2 p^2}{p^+}\\\\
     1& \frac{p^1}{2 \bar p}       & 0 & 1 & 0\\\\
      2&  \frac{p^2}{2 \bar p}       & 0 & 0 & 1
}. \label{vec_l}
\eeq
Comparing Eq.(\ref{vec_l}) with Eq.(\ref{vecl}), 
we see that in contrast to the helicity choice, the expressions of  the light-front boost are almost identical for both massive and massless particles. The only difference is that $m$ in the massive case is replaced by $2 \bar p$ in the massless case.\\

\end{enumerate}

\underline{Remark A} Using  the explicit vector representation in Eq.(\ref{vec_h}) and (\ref{vec_l}) , one finds that for massless particles, the spin vector defined by  the Wick helicity boost points in the direction of the $3$-momentum $\mathbf{p}$, and the spin vector defined by the light-front boost points in the light-front $3$-momentum $(p^+, p^\perp)$. This is exactly what we found in the massive case. Therefore, we conclude that $\Lambda_h$ and $\Lambda_L$ can be used to define spin states for massive and massless particles for any  momentum $p$. 

\underline{Remark B} It is worth mentioning that the massless spin-$1$ representation defined by the light-front boost preserves the light-cone gauge condition \\$A^+=0$ under Lorentz transformations \footnote{The explicit light-front spin representations for spin-$1$ and spin-$\frac{1}{2}$ particles are given in Appendix B; the preservation of the $A^+=0$ condition is demonstrated in Eq. (\ref{124}).}. Thus, in contrast to  other choices of Lorentz transformations where gauge conditions are generally not preserved, one can always choose the 
$A^+=0$ gauge condition in all Lorentz frames. We also emphasize  that the light-front Lorentz transformations are \textit{kinematical} and leave the $x^+=0$ plane invariant, unlike the canonical or Wick helicity boosts.

\section{Conservation of Angular Momentum : A Property of the Light-front Lorentz Transformation}
In the last section, we have defined  different choices of Lorentz transformations, which up to this point merely look like a preference of choices. However, as we will discover in this section, the light-front choice is advantageous and unique in that it provides frame-independent angular momentum conservation rules. 

We shall first study the action of different of Lorentz transformations on the spin.  

 \subsection{Spin of particles in motion -- Why is light-front  special?}

Let us start with a spin state pointed in the $z$-direction in a massive particle's rest frame:
\begin{align}
\ev{S^i}(\mathring p)=\frac{\bra{\mathring p; \lambda=s^3} S^i \ket{\mathring p; \lambda=s^3}}{\braket{\mathring p; \lambda|\mathring p; \lambda}} ={\lambda} \begin{bmatrix}
0 \\
0\\
1
\end{bmatrix}, \;\; i=1,2,3.
\end{align}
We then wish to find the spin of the particle when it is motion
\begin{align}
\ev{S^i}(p)=\frac{\bra{p; \lambda} S^i \ket{p; \lambda}}{\braket{p; \lambda|p; \lambda}} \label{29}
\end{align}
using different choices of Lorentz transformations. Note that  $\ev{S^{i}}(p)$ in Eq.(\ref{29}) are the \textit{expectation values} of spin operators on moving states. The $3$-vector formed by $\ev{S^i}(p)$  should not be confused with the spatial components of the covariant\footnote{A detailed discussion on the definition of the covariant spin vector can be found in Appendix C.1.} spin $4$-vector defined by the expectation value of the Pauli-Lubanski operator.\\\\
The calculation  can be done with the definitions of Lorentz transformations in Eq. (\ref{lac}), (\ref{lah}), (\ref{lal}) and the commutation relation 
\beq
[M^{\alpha\beta},  M^{\mu\nu}] = i\left( g^{\alpha\nu}  M^{\beta\mu} + g^{\beta\mu}  M^{\alpha\nu} - g^{\alpha\mu}  M^{\beta\nu} - g^{\beta\nu}M^{\alpha\mu}\right).
\eeq
 Since the Lorentz algebra is representation independent, one can pick any representation to do the computation without loss of generality. For example, let us consider the spin$-1$ representation and the matrices given in Eq. (\ref{vecc})  (\ref{vech}) and (\ref{vecl}). We then obtain the following results\footnote{We have verified that the analogous calculation for massless particles using Eq. (\ref{vec_h}) and (\ref{vec_l}) yields the same $\ev{S^i}_h(p)$ and $\ev{S^i}_L(p)$, whereas $\ev{S^i}_c(p)$ is well-defined only for massive particles.}:
\\
\begin{enumerate}
\item Canonical spin
\begin{align}
\ev{S^i}_c(p)=\frac{_c{\bra{p; \lambda}} S^i \ket{p; \lambda}_c}{\indices{_c}{\braket{p; \lambda|p; \lambda}_c}} =\frac{\lambda}{m(m+p^0)} \begin{bmatrix}
-p^1 p^3 \\
-p^2 p^3\\
m(m+p^0)+ {\abs{p^\perp}}^2
\end{bmatrix}.
\end{align}
This shows the canonical spin is generally not aligned along the direction of motion, as we have already seen in Fig.\ref{spins_a}.

\item Wick helicity spin
\begin{align}
\ev{S^i}_h(p)=\frac{\indices{_h}{\bra{p; \lambda}} S^i \ket{p; \lambda}_h}{\indices{_h}{\braket{p; \lambda|p; \lambda}_h}} =\frac{\lambda}{\abs{\mathbf p}} \begin{bmatrix}
p^1 \\
p^2\\
p^3
\end{bmatrix}.
\end{align}
The helicity spin points along the $3$-momentum direction.   For example,  a photon moving in the $x$-direction has two states polarized along the $x$-axis; thus,  $\ev{S^{i=1}}_h(p)=\pm 1$  in the helicity spin definition, as illustrated in Fig.  \ref{spins_b}. Note however that  $\ev{S^{i=3}}_h(p)= \lambda \frac{p^3}{\abs{\mathbf p}} \neq \lambda$, and therefore one cannot identify the $z$-component of spin with the helicty $\lambda$ for particles in motion. 

\item Light-front spin
\begin{align}
\ev{S^i}_L(p)=\frac{\indices{_L}{\bra{p; \lambda}} S^i \ket{p; \lambda}_L}{\indices{_L}{\braket{p; \lambda|p; \lambda}_L}} =\lambda \begin{bmatrix}
\frac{p^1}{p^+} \\[5pt]
\frac{p^2}{p^+} \\[5pt]
1
\end{bmatrix}. \label{LF_s}
\end{align}\\
\end{enumerate}

\underline{Remark C} Note that for all three choices of Lorentz transformation,  the norm of the spin expectation value  for moving particles, $\ev{S^i}(p)$,  is not conserved, in contrast to the nonrelativistic  case where spin is a $3$-vector with unit norm. This is a consequence of the non-unitarity of the boost operation. Nevertheless, using the light-front definition, we find $\ev{S^{i=3}}_L(p)= \lambda= s^3$ and the spin projection along the \textit{light-front} direction $\hat z= \hat 3$ in any Lorentz frame is always the same as in the rest frame. Thus, $s^3$ is an invariant under the light-front choice of Lorentz transformation.

\underline{Remark D} In the non-relativistic regime where ${\abs{\mathbf p}}^2 \ll m^2$,  $\ev{S^i}_c(p)$  and $\ev{S^i}_L(p)$ reduces to the usual spin definition which is frame independent:
\begin{align}
\ev{S^i}_c(p)=\ev{S^i}_L(p) \underset{{\abs{\mathbf p}}^2 \ll m^2}{\rightarrow} \; \lambda 
\begin{bmatrix}
0 \\[5pt]
0\\[5pt]
1
\end{bmatrix}
= \ev{S^i}(\mathring p).
\end{align}

\underline{Remark E} On the other hand, in a reference frame where the observer moves with infinite momentum in the negative $z$-direction and $ p^3 \approx \abs{\mathbf p}$, the Wick helicity spin is 
\begin{align}
\ev{S^i}_h(p) \underset{p^3 \approx \abs{\mathbf p}}{\rightarrow} \; \lambda \begin{bmatrix}
\frac{p^1}{p^3} \\[5pt]
\frac{p^2}{p^3} \\[5pt]
1
\end{bmatrix}. \label{IMF_s}
\end{align}
Applying the usual identification of $p^3$ in the  infinite momentum frame (IMF) with $p^+$ in the front  form \cite{PhysRev.150.1313} \cite{Brodsky:1973kb},  Eq. (\ref{IMF_s}) becomes the same as $\ev{S^i}_L(p)$  in Eq. (\ref{LF_s}). Therefore, the Wick helicity spin in the IMF is the same as the light-front spin, and the $z$-component of the Wick helicity spin remains invariant in the IMF. This is one of the ways that one can see the correspondence between the IMF and the front form.

In summary, the light-front spin is powerful because: (i) it is applicable to both nonrelativistic and relativistic regimes; (ii) one does not need the IMF  to show that the spin along the $z$-direction is preserved. In the front form, $\ev{S^{i=3}}_L(p)=\lambda=s^3$ is true in all Lorentz frames. The invariance  of $s^3$ in the front form provides a great advantage for the angular momentum sum rules for composite systems, which we will explore in Section 4.

\subsection{Invariance of light-front spin for elementary particles}
In this section, we shall give a formal proof of the invariance of spin for  elementary particles  under the light-front Lorentz transformation.

Let us start by defining an operator ${S^3_L}(p)$, which when acting on a light-front state $\ket{p;\lambda}_L$ gives $s^3$ -- the rest frame spin projection along the $z$-axis:
\beq
{S^3_L}(p) \ket{p; \lambda}_L = s^3  \ket{p; \lambda}_L.  \label{s3p}
\eeq
Using the definition of the light-front spin state
\beq
\ket{p; \lambda}_L=\Lambda_L(\mathring{p} \rightarrow p)  \ket{\mathring{p}; \lambda} \label{s3_1}
\eeq
and 
\beq
J^3\ket{\mathring{p}; \lambda}=s^3 \ket{\mathring{p}; \lambda},
\eeq
one deduces
\beq
{S^3_L}(p) = \Lambda_L(\mathring{p} \rightarrow p) \; J^3  \; {\Lambda^{-1}_L}(\mathring{p} \rightarrow p). 
\eeq

Furthermore, we can express ${S^3_L}(p)$  in terms of the \Poincare generators as\footnote{One can repeat the same calculation for massless particles and obtain the same expression. }
\begin{align}
{S^3_L}(p)&= J^3- \frac{P^1}{P^+} M^{+2} +\frac{P^2}{P^+} M^{+1} \label{45}\\
&=J^3- \frac{P^1}{P^+} S^{+2} +\frac{P^2}{P^+} S^{+1} - L^3_L(p), \label{til_l}
\end{align}
where 
\beq
 L^3_L(p)=\frac{P^1}{P^+} L^{+2} -\frac{P^2}{P^+} L^{+1}, \label{42}
\eeq
and $S^{\mu\nu}$ and $L^{\mu\nu}$ are the spin and the orbital part of the Lorentz generators, respectively.

We can now compute the total angular momentum for a moving particle 
\begin{align}
\ev{J^3}_L(p)= 
\frac{\indices{_L}{\bra{p; \lambda}} J^3 \ket{p; \lambda}_L}{\indices{_L}{\braket{p; \lambda|p; \lambda}_L}}. \label{s_l} 
\end{align}
Rewriting Eq.(\ref{s_l}) using Eq.(\ref{til_l}) and  (\ref{s3_1}), we have 
\begin{align}
\ev{J^3}_L(p)= \ev{S^3_L(p)} + \; \frac{\indices{_L}{\bra{p; \lambda}}\frac{P^1}{P^+} S^{+2} -\frac{P^2}{P^+} S^{+1}\ket{p; \lambda}_L}{\indices{_L}{\braket{p; \lambda|p; \lambda}_L}} + \; \frac{\indices{_L}\bra{p; \lambda} L^3_L(p)\ket{p; \lambda}_L}{\indices{_L}{\braket{p; \lambda|p; \lambda}_L}} \label{j3l}
\end{align}
We will show below that the last two terms in fact vanish, and then since $\ev{J^3}_L(p)= \ev{S^3_L(p)}= s^3$, we  prove the invariance of spin along the $z$-direction under light-front Lorentz transformations.  \\ 
 
In the second term, the matrix
\begin{align}
\bra{p; \lambda}S^{+\perp}\ket{p; \lambda}_L 
&= \bra{\mathring p; \lambda}  {\Lambda^{-1}_L}(\mathring{p} \rightarrow p) \;S^{+\perp}  \; \Lambda_L (\mathring{p} \rightarrow p) \ket{\mathring p; \lambda}\\
&= \bra{\mathring p; \lambda}  e^{i \frac{S^{+-}\omega}{2}}\; S^{+\perp}\; e^{-i \frac{S^{+-}\omega}{2}}\ket{\mathring p; \lambda}, \; \text{with\;}e^\omega= \frac{m}{p^+}\\
&= e^{\omega} \bra{\mathring p; \lambda}  S^{+\perp}\ket{\mathring p; \lambda}\\
&=0. \label{sv}
\end{align}
The second line is obtained by using the definition of light-front boost in Eq. (\ref{lal}), along with  $\left[ S^{+ 1},S^{+ 2} \right]=0 $, and the fact that the spin and  orbital Lorentz generators commute $\left[S^{\mu\nu}, L^{\alpha\beta} \right]=0$. The third line is due to the property that in the front form,  the kinematical generators are invariant up to a scaling under a longitudinal boost, and 
\beq
\displaystyle e^{i \frac{S^{+-}\omega}{2}}\; S^{+\perp}\; e^{-i \frac{S^{+-}\omega}{2}}= e^\omega S^{+\perp}.
\eeq
To obtain the last line, recall that the state at rest $\ket{\mathring p; \lambda}$ is an eigenstate defined by the eigenvalues of $J^3$ as in Eq. (\ref{so3}).  Therefore the expectation values for all other Lorentz generators on $\ket{\mathring p; \lambda}$ vanish, and hence the last equality.   \\
   
The last term can be simplified in the following way: 
\begin{align}
\ev{L^3_L(p)}&={_L}{\bra{p; \lambda}} L^3_L(p)\ket{p; \lambda}_L\\[4pt]
&=  {_L}{\bra{p; \lambda}} 
i(p^1 \frac{\partial}{\partial p_2} - p^2 \frac{\partial}{\partial p_1} )\ket{p; \lambda}_L \label{51}\\[4pt]
&={_L}{\bra{p; \lambda}} L^3\ket{p; \lambda}_L  \\ 
&\equiv\ev{L^3}_L(p). \label{elem}
\end{align}
The second line is obtained by using the explicit form of $L^3_L(p)$ in Eq.(\ref{42}) and the fact  any function of generator $P^\mu$ on the momentum eigenstate satisfies \\$f(P^\mu)\ket{p}= f(p^\mu)\ket{p}$. The third equality is then obvious by noting that the momentum-space representation for the orbital angular momentum operator is  $\displaystyle L^{\mu\nu}= i (p^\mu \frac{\partial}{\partial p_\nu} - p^\nu \frac{\partial}{\partial p_\mu} )$.\\

Combining Eq. (\ref{sv}) and (\ref{elem}), We obtain a simplified expression for Eq.(\ref{j3l}): 
\begin{align}
\ev{J^3}_L(p)&= \ev{S^3_L(p)} + \ev{L^3}_L(p). \label{59}
\end{align}
Comparing this  with usual expression for angular momentum conservation :
\begin{align}
\ev{J^3}_L(p)&= \frac{\indices{_L}{\bra{p; \lambda}} J^3 \ket{p; \lambda}_L}{{_L}{\braket{p; \lambda|p; \lambda}_L}}\\
 &= \frac{\indices{_L}{\bra{p; \lambda}} \left( S^3 + L^3 \right) \ket{p; \lambda}_L}{{_L}{\braket{p; \lambda|p; \lambda}_L}}\\[5pt]
&= \ev{S^3}_L(p) + \ev{L^3}_L(p),
\label{61}
\end{align}
we deduce 
\beq
\ev{S^3}_L(p)=\ev{S^3_L(p)}= s^3.
\label{62_1}
\eeq

Let us discuss the interpretation of $\ev{L^3}_L(p)$ for elementary particles without internal structure. An elementary particle with fixed momentum is described by a plane wave which is  spread all over the space and carries no orbital angular momentum around a fixed point. It has also been shown with detailed wavepacket analysis in \cite{Bakker:2004ib} and \cite{Jaffe:1989jz} that this term is regulated and contains no infinities and only depends on the particle's motion around a fixed center; thus, this term has no applicability to the internal spin. We can thus neglect this term and only discuss the intrinsic spin angular momentum for elementary particles. Thus, we find
\beq
\ev{J^3}_L(p)= \ev{S^3}_L(p)= s^3=\lambda.
\label{62}
\eeq

This proves the invariance of spin  for elementary particles -- in any Lorentz frame obtained by a light-front Lorentz transformation from the particle's rest frame, the expectation value of the spin-projection operator along the $z$-direction is the same as in the particle's rest frame. \\

\underline{Remark F} In the literature, people often say ``light-front spin" or ``light-front helicity" is invariant. The accurate statement should be: the spin along the $z$-direction defined by the light-front Lorentz transformation is preserved   because $\ev{J^3}_L(p)=  s^3$ for all momentum $p$. Furthermore, since the helicity $\lambda$ is equal to $s^3$ by definition, spin and helicity can thus be used interchangeably in the front form. Similarly, in the operator level, since $S^3_L(p)$  also gives the Lorentz-invariant $z$-component of spin  for particles as in Eq.(\ref{62}), it is often referred to as the ``light-front spin operator" or ``light-front helicity operator" \cite{Leutwyler:1977vy} \cite{Brodsky:1997de} \cite{Harindranath:2013goa}.  In addition to $S^3_L(p)$, one can also define the transverse light-front spin operators $S^\perp_L(p)$ in analogy to Eq.(\ref{s3p}). The relation between the light-front spin operators $S^i_L(p)$ and the Pauli-Lubanski vector $W^\mu$ is given in Appendix C.2.


\underline{Remark G} One may wonder whether the invariance of the $z$-projection of spin  also occurs in the Wick helicity boost, which is defined similarly to the light-front boost. The answer is yes, but only in the IMF limit\cite{Soper:1972xc}.

To see this, we first construct $S^3_h(p)$  for the Wick helicity boost analogous to the light-front  ${S^3_L}(p)$:
\beq
S^3_h(p) \ket{p; \lambda}_h = \lambda  \ket{p; \lambda}_h.  
\eeq
This ``Wick helicity spin operator" satisfies
\beq
S^3_h(p) = \Lambda_h(\mathring{p} \rightarrow p) \; J^3  \; {\Lambda^{-1}_h}(\mathring{p} \rightarrow p).
\eeq
An explicit calculation gives
\begin{align}
S^3_h(p)&= \frac{J^i P^i}{\abs{\mathbf P}}, \label{64}
\end{align}
which is exactly the ordinary helicity operator. 

Even though $s^3=\lambda$ in the standard reference frame, in general $\ev {S^3} _h (p) \neq \ev{S^3_h(p)}=\lambda$ for an arbitrary momentum $p$. Nevertheless, in the IMF where $ p^1, p^2 \ll  p^3 \approx \abs{\mathbf p}$,  
\begin{align}
S^3_h(p) \underset{p^3 \approx \abs{\mathbf p}}{\rightarrow} \;J^3+ \frac{P^1}{P^3} M^{23} +\frac{P^2}{P^3} M^{31}. \label{66}
\end{align}
Identifying $p^3$ in the IMF with $p^+$ in the front form, 
Eq.(\ref{66}) becomes identical to $S^3_L(p)$ in Eq.(\ref{45}). We see that in the IMF the Wick helicity spin operator $S^3_h(p)$  is the same as the light-front spin operator $S^3_L(p)$ in the front form.  Alternatively, one may take $ p^1, p^2 \rightarrow 0$ in the IMF  and find  $S^3_h(p)  \underset{p^1, p^2 \approx 0}{\rightarrow} J^3$. Both ways give
\begin{align}
\ev{S^3_h(p)} = \ev{{S^3_L}(p)}=\lambda=s^3
\end{align}
in the IMF. 

This explains why the $z$-projection of spin is preserved in the IMF, which we have already seen in Eq.(\ref{IMF_s}).

\subsection{Conservation of $j^3$ for composite systems in the front form}
In this section, we shall generalize the proof to  composite systems and show that the $z$-component of the total angular momentum  is conserved for any bound state in the front form.

Bound states in the front form are defined at one instant of light-front time $x^+=0$. As we will see, light-front bound state wavefunctions are in fact \Poincare invariant, in contrast to instant-form wavefunctions defined at $x^0=0$.   A bound state with momentum $p$  has the following light-front Fock state decomposition \cite{Brodsky:2000ii}
\beq
\ket{p;j^3}_L= \sum_n \int \; [\dd x ] [\dd^2 k^\perp ] \;\psi_n(x_a, k^\perp_a, s^3_a) \ket{n; p_a; s^3_a }_L,\;\; \forall\; s^3_a,
\label{qed}
\eeq
with the Lorentz invariant integral measure
\begin{align}
&\left[\dd x \right]=   \prod^n_{a=1}   \frac{\dd x_a}{  \sqrt {x_a}}  \delta (1- \sum^n_{a=1} x_a) \;\;
 &\left[\dd^2 k^\perp \right]= {16 \pi^3} \prod^n_{a=1}  \frac{\dd^2 k^\perp_a}{16 \pi^3} \;   \delta^2 (\sum^n_{a=1} k^\perp_a).
\end{align}
The total angular quantum number $j^3$ can be  defined in the standard reference frame in which the bound state is at rest, analogous to the case of massive elementary particles in Section 2. $\ket{n; p_a; s^3_a }_L$ denotes the $n$-particle Fock state; $a$  labels the $n$ constituents; $s^3_a$ is  the $z$-projection of the light-front spin for each of the constituents, which we have proved to be Lorentz invariant in the previous section.
 The light-front $3$-momentum of  constituent $a$ is given by
\begin{align}
&p^+_a= x_a p^+   &p^\perp_a= x_a p^\perp + k^\perp_a.   
\end{align}
The light-front energy  of a constituent is given by $\displaystyle p^-_a= \frac{{(p^\perp_a)}^2+ m^2_a}{p^+_a}$, where $m_a$ is the mass of the constituent. For a bound state, the light-front energy is given by $\displaystyle p^-= \frac{{(p^\perp)}^2+ m^2}{p^+}$, where $m$ is the mass of the bound state. Note that  $\displaystyle p^- \neq \sum_{a=1}^n p^-_a $ because in light-front time-ordered perturbation theory, particles are always on their mass shell, but off the energy conservation shell.  Thus, it is sufficient to specify a bound state with the total light-front $3$-momentum $(p^+, p^\perp)$ of the bound state  together with the internal variables $(x_a, k^\perp_a)$.

It can be readily checked that $(x_a, k^\perp_a)$ are in fact \Poincare invariant, despite that $(p^+_a, p^\perp_a)$ transform covariantly under the light-front Lorentz transformation defined in Eq.(\ref{vecl}).
Since $s^3_a$ is Lorentz invariant,   the light-front wavefunction(LFWF) $\psi_n(x_a, k^\perp_a, s^3_a)$, which describes the internal structure of a bound state, is indeed  independent of the observer's Lorentz frame as desired.

We compute the total angular momentum along $z$-direction for each of the $n$-particle Fock state $ \ket{n; p_a; s^3_a }$ at arbitrary momentum $p$:  
\begin{align}
 \ev{J^3}_L(p)&= \frac{{_L}{\bra{n; p_a; s^3_a }} \;J^3\; \ket{n; p_a; s^3_a }_L}{{_L}{\braket{n; p_a; s^3_a|n; p_a; s^3_a }_L}}\\[5pt]
 &= \sum_{a=1}^n  \frac{{_L}{\bra{n; p_a; s^3_a }} \;(S^3_a + L^3_a)\; \ket{n; p_a; s^3_a }_L}{{_L}{\braket{n; p_a; s^3_a|n; p_a; s^3_a }_L}}\\[5pt]
&=  \sum_{a=1}^n \;\; s^3_a   
+  \frac{{_L}{\bra{n; p_a; s^3_a }} \; L^3_a\; \ket{n; p_a; s^3_a }_L}{{_L}{\braket{n; p_a; s^3_a|n; p_a; s^3_a }_L}}.\label{74}
\end{align}
We have used conservation of spin for elementary constituents in Eq. (\ref{62_1}) to obtain the last equality.

We shall show below that not only $s^3_a$ is Lorentz invariant, but the orbital angular momentum along the $z$-direction is also independent of the observer's Lorentz frame. We rewrite the orbital term with  the total transverse momentum $p^\perp$ of the bound state and the $n-1$ independent  internal transverse momentum $k^\perp$ \cite{Harindranath:1998ve}: 
\begin{align}
&\sum_{a=1}^n  {_L}{\bra{n; p_a; s^3_a }} \;   L^3_a \; \ket{n; p_a; s^3_a }_L \\
= &\sum_{a=1}^n  
{_L}{\bra{n; p_a; s^3_a }} \;  i (p^2_a \frac{\partial}{\partial p^1_a} - p^1_a \frac{\partial}{\partial p^2_a} )\;  \ket{n; p_a; s^3_a }_L\\
= &{_L}{\bra{n; p_a; s^3_a }}  \;  i (p^2\frac{\partial}{\partial p^1} - p^1\frac{\partial}{\partial p^2} ) \; \ket{n; p_a; s^3_a }_L \nonumber\\
&\;+ \sum_{a=1}^{n-1} {_L}{\bra{n; p_a; s^3_a }} \;   i (k^2_a \frac{\partial}{\partial k^1_a} - k^1_a \frac{\partial}{\partial k^2_a} ) \; \ket{n; p_a; s^3_a }_L.  
\label{73}
\end{align}
The first term in Eq.(\ref{73}) corresponds to the orbital angular momentum due to total momentum of the composite system and is thus neglected due to its irrelevance to the internal structure. The  second term depends on the frame-independent internal transverse momentum $k^\perp_a$, and thus it gives the Lorentz invariant internal orbital angular momentum $l^3_a$ for the constituents.  Therefore, we deduce
\begin{align}
&\sum_{a=1}^n  {_L}{\bra{n; p_a; s^3_a }}\;   L^3_a \; \ket{n; p_a; s^3_a }_L
=   \sum_{a=1}^{n-1} \; l^3_a \; {_L}{\bra{n; p_a; s^3_a }}    \ket{n; p_a; s^3_a }_L. \label{78}
\end{align}
Inserting Eq.(\ref{78}) into Eq.(\ref{74}), we find that the $z$-projection of the total angular momentum ${\ev {J^3}}_L(p)$ satisfies 
\begin{align}
{\ev {J^3}}_L(p)=j^3= \sum_{a=1}^n \; s^3_a  + \sum_{a=1}^{n-1} \; l^3_a \label{j3con}.
\end{align}
 
Since $s^3_a$ and $l^3_a$ are Lorentz invariant, ${\ev {J^3}}_L(p)= j^3$ must also be Lorentz invariant. This thus completes the proof of the Lorentz invariance of the angular momentum conservation law. 

In summary, we have proved that, in the front form the internal angular quantum number $j^3$ is frame independent and is determined only by  the internal structure of the composite system. This is an important consequence that boosts are \textit{kinematical} in the front form\cite{RevModPhys.21.392}. Due to its Lorentz invariance, this conservation law is rigorous and can be applied to any composite system at all momentum scales.

\subsection{Alternative proof of $j^3$ conservation using Lorentz algebra in the front form}
The conservation of angular momentum for a composite system can be understood from another perspective using  the Lorentz algebra in the front form.  

In the front form,   the action of the set of kinematical generators \\ $(M^{+-},P^+, M^{12})$ on $M^{+\perp}$  is identical to its action on  $P^\perp$  up to a scaling. To see this, we compare
\begin{align}
e^{iM^{12}\phi}\;&M^{+\perp} \;e^{-iM^{12}\phi}= cos\phi \,M^{+\perp} - sin\phi\, \varepsilon^{\perp \perp '} M^{+\perp '} \\
e^{i \frac{M^{+-}\omega}{2}}\;
&M^{+\perp}\; e^{-i \frac{M^{+-}\omega}{2}}= e^{\omega} M^{+\perp } \\
e^{i P^+ x^-}\;
&M^{+\perp}\; e^{-i P^+ x^-}=  M^{+\perp },
\end{align}
with
\begin{align}
e^{iM^{12}\phi}\;&P^{\perp}\; e^{-iM^{12}\phi}= cos\phi \,P^\perp - sin\phi\, \varepsilon^{\perp \perp '} P^{\perp '} \\
e^{i \frac{M^{+-}\omega}{2}}\;
&P^{\perp}\; e^{-i \frac{M^{+-}\omega}{2}}=  P^{\perp } \\
e^{i P^+ x^-}\;
&P^{\perp}\; e^{-i P^+ x^-}=  P^{\perp },
\end{align}
where $\varepsilon^{12}= -\varepsilon^{21}=1$, and $\perp, \perp'=1,2$.

The relations suggest that, when evaluating  $\ev{J^3}_L(p)$ for a light-front spin state $\ket{p;j}_L$, which depends on the \textit{kinematical} Lorentz generators $M^{+-}$ and $M^{+\perp}$, the action of $M^{+\perp}$ on the state should be the same as the transverse translation generator $P^{\perp}$. 

Since a translation does not change a particle's angular momentum  and the boost in the $z$-direction $M^{+-}$ commutes with $J^3$, we  deduce that the $z$-projection of the angular momentum is preserved in  light-front boosts, and $\ev{J^3}_L(p)= j^3$.

 This result is again a reflection that boosts in the front form are kinematical, in contrast to the instant form. In fact, this statement can be generalized to all kinematical transformations in the front form -- any transformation generated by the kinematical subgroup of the \Poincare group leaves $j^3$ invariant in the front form.

\section{Selection Rule for  Orbital Angular \\Momentum in the Front Form}
In this section, we  apply the angular momentum conservation law in the light-front formulation derived   in Section 3.3 to present an explanation for the selection rule of the orbital angular momentum observed in  \cite{Cruz-Santiago:2015dla}　\cite{Brodsky:2000ii}: in the $n$-th order perturbative expansion of a renormalizable  theory, the change of orbital angular momentum between the initial and final states in the front form is constrained by $|\Delta\, l^3|\leq n$.

Recall that in \textit{nonrelativistic} quantum mechanics, a state's orbital angular momentum is changed by  $1$ unit when it is acted on by  the  transverse \textit{circular} momentum  operator $P^\mathrm R \equiv P^1+ i P^2$ and $P^\mathrm L \equiv  P^1- i P^2$: 
\begin{align}
P^{\mathrm R} \ket{l^3} \propto \ket{l^3+ 1} &&
P^{\mathrm L} \ket{l^3} \propto \ket{l^3- 1}
\end{align}
This is true because
\begin{align}
[P^{\mathrm R}, J^3]= - P^{\mathrm R} && [P^{\mathrm L}, J^3]=  P^{\mathrm L} 
\end{align}
and
\begin{align}
&J^3 \;\,P^{\mathrm R} \ket{l^3}=(l^3+ 1)\;\,P^{\mathrm R}\ket{l^3}  \; \Rightarrow P^{\mathrm R} \ket{l^3} \;\propto \ket{l^3+ 1}\\
&J^3 \;\,P^{\mathrm L} \ket{l^3}=(l^3- 1)\;\,P^{\mathrm L}\ket{l^3}  \; \Rightarrow P^{\mathrm L} \ket{l^3} \;\propto \ket{l^3- 1}.
\end{align}
In general,
\begin{align}
(P^{\mathrm R})^n \;\ket{l^3} \propto \ket{l^3+ n} &&
(P^{\mathrm L})^n\; \ket{l^3} \propto \ket{l^3- n}.
\end{align}

It follows that, an interaction $H_I$ proportional to  $n$ powers of the transverse momentum $P^\perp=(P^1, P^2)$ can change a state's orbital angular momentum  at most by $n$:
\beq
\bra{p', l'} H_I \ket{p,l} =0   , \text{\;\;\; for\;\;\;}  |\Delta\, l^3|\geq n .
\label{86}
\eeq \\

For  \textit{relativistic} quantum field theories in the instant form,  the angular momentum in the $z$-direction generally changes under Lorentz transformation, and thus the above selection rule  cannot be easily applied to relativistic systems in the instant form since $l^3$ is not Lorentz invariant. Nonetheless, in the front form, the  angular momentum conservation law  
\begin{align}
j^3= \sum_{a=1}^n \; s^3_a  + \sum_{a=1}^{n-1} \; l^3_a, 
\end{align}
which we derived in Eq.(\ref{j3con}) is \textit{frame independent}. The  quantum numbers $(s^3,l^3,j^3)$ are in fact invariant under the light-front Lorentz transformations, and only depend on the internal angular structure of  particles. Therefore, we can readily apply the  orbital angular momentum selection rule in Eq.(\ref{86}) to constrain the change of the orbital angular quantum number in interactions.

Specifically, in all renormalizable theories, since the interaction vertex $H_I$ only contains  at most one power of $P^\perp$,  the change of orbital angular momentum is constrained by 
\begin{align}
|\Delta\, l^3|\leq 1 
\end{align}
 at every vertex.  

This selection rule can be used to eliminate certain interaction vertices in QED and QCD. For example, in $e^- \rightarrow e^-\gamma$ scattering, if the incoming electron line has $ s^3_{in}=-\frac{1}{2}$, then it is not possible to have outgoing lines  with $ s^3_{out}=s^3_e+ s^3_\gamma= +\frac{1}{2} +1=+\frac{3}{2}$ because $\Delta\, l^3= -\frac{3}{2}$ in this case. One can also use the spin representations given in Appendix B to explicitly verify $V(-\frac{1}{2} \rightarrow +\frac{1}{2} +1 )=0$. Similarly, in QCD, the  $3$-gluon interaction vertex $V(- \rightarrow ++; \;\Delta\, l^3= -2)$  and the $4$-gluon  vertices $V(- \rightarrow +++;\; \Delta\, l^3= -4)$, $ V(- \rightarrow -++;\; \Delta\, l^3= -2)$ and $V(- \rightarrow ---;\;\Delta\, l^3= 2)$ all vanish by the same argument.

Furthermore, in the $n$-th order perturbative expansion, the change of between initial and final state orbital angular momentum  is constrained by 
\begin{align}
 |\Delta\, l^3|\leq n.
\end{align}
This explains the vanishing  amplitude $\mathcal{M}(+,+,...,+)$\footnote{In this convention, all momenta are assumed to be outgoing. } at tree level\;\cite{Cruz-Santiago:2015dla} \cite{Dixon:1996wi} :\\ In  $2\rightarrow n$ gluon scattering , the amplitude  $\mathcal{M}(+,+,...,+)=\mathcal{M}(-- \rightarrow +...+)$ has  $\,\Delta\, s^3= n+2$. Conservation of angular momentum in the $z$-direction then gives  $\,\Delta\, l^3= -(n+2)$.  At tree level, since there are only at most $n$ triple gluon vertices in this process and $|\Delta\, l^3|\leq n$,  $\mathcal{M}(+,+,...,+)$ must then vanish due to violation of the selection rule.

\section{Discussion} 
In this paper we have proved that  the $z$-component of the total angular momentum of any system is invariant  under Lorentz transformations in the front form. In particular, we have demonstrated that for a bound state, the internal angular quantum numbers $(j^3, s^3, l^3)$ which appear in the light-front wavefunctions are  independent of the observer's Lorentz frame. In contrast to \cite{Soper:1972xc} \cite{Leutwyler:1977vy} \cite{Li:2015hew} \cite{Brodsky:1997de}  \cite{Harindranath:2013goa}, where  $j^3$ is understood as the eigenvalue of the light-front helicity operator, which is not a charge operator of the Lorentz symmetry, we showed explicitly that in fact, $j^3$ also corresponds to the expectation value of the actual total angular momentum operator. This provides an explanation to the conservation of $j^3$, which has been implicitly assumed in  \cite{Cruz-Santiago:2015dla} \cite{Brodsky:2000ii} \cite{PhysRevD.24.2848}.

The conservation of $j^3$ is an important consequence of the fact that the light-front boosts are kinematical, which leave the $x^+ = 0 $ quantization plane invariant. These quantum numbers can be applied to particles in the intermediate states which are off-shell in the light-front energy $p^-$. Moreover, $j^3$  is conserved
for any intermediate states even though they are off-shell.  In addition, we have shown that the $A^+ = 0$ light-front gauge condition is preserved under Lorentz transformations in the front form. Thus, one can consistently use  light-front gauge in all Lorentz frames, avoiding the redundant gauge degrees of freedom characteristic of covariant gauges.

We applied the angular momentum conservation law and found an upper bound for the change of orbital angular momentum between initial and final states in scattering processes -- in a renormalizable theory,  $|\Delta\, l^3|\leq 1$ at every vertex and  $|\Delta\, l^3|\leq n$ in the $n-$th order perturbative expansion. We also showed explicitly that this selection rule can be used to eliminate certain interaction vertices in QED and QCD scattering processes.

In order to understand the specific features of the front form, we analyzed the spin states defined by different choices of Lorentz transformations and found that:
(i) in the non-relativistic limit, light-front spin is identical to the canonical spin,  and  (ii) in the infinite momentum frame(IMF) limit, the Wick helicity spin reduces to the light-front spin, which explains the conservation of helicity in the IMF. 

Thus, we conclude that the light-front spin is suitable for describing the spin structure of particles at all momentum scales.

\section*{Acknowledgements}
The authors would like to thank Peter Lowdon, Michael Peskin, Cedric Lorce, Qin Chang, and Andrew McLeod for useful discussions. This work was supported by the US Department of Energy under contract DE--AC02--76SF00515.

\cleardoublepage

\begin{appendices}

\section{Notation}
\subsection{Glossary of symbols}
Throughout the paper, we use uppercase letters to denote operators, and lowercase letters to denote the value of the operator acting on some states.  

The $SU(2)$ spin operators are $S^i$, for $i=1,2,3$. 

The average spin $s^i$ of a particle at rest is 
\begin{align}
 s^i =
 \begin{bmatrix}
s^1\\
s^2\\
s^3
\end{bmatrix}= \frac{1}{s}\,\bra{ s^i}  \begin{bmatrix}
S^1\\
S^2\\
S^3
\end{bmatrix}  \ket{ s^i}.
\end{align}

The average spin of a particle in motion is denoted by 
\begin{align}
\ev{S^i}(p)=\frac{\bra{p; \lambda} S^i \ket{p; \lambda}}{\braket{p; \lambda|p; \lambda}}.
\end{align}

A relativistic spin state $\ket{p; \lambda}$ defined by a Lorentz transformation $\Lambda$ from the standard reference frames in which spin is labeled along the $z$-direction by $\lambda=s^3$. For massive particles, the standard reference frame is the rest frame, and  
\begin{align}
\ket{p; \lambda} & \equiv \Lambda(\mathring{p} \rightarrow p)  \ket{\mathring{p}; s^3= \lambda}.
\end{align}
For massless particles, the standard reference frame is in which the particle moves along the $z$-direction so that the helicity coincides with the $z$-component of the spin
\begin{align}
\ket{p; \lambda} &\equiv \Lambda(\bar{p} \rightarrow p)  \ket{\bar{p}; s^3= \lambda}.
\end{align}

There are three choices of Lorentz transformations in general use: canonical $\Lambda_c$, helicity $\Lambda_h$ and light-front $\Lambda_L$.

Light-front spin operator or light-front helicity operators $S^3_L(p)$ are defined by 
\beq
{S^3_L}(p) \ket{p; \lambda}_L = s^3  \ket{p; \lambda}_L,
\eeq
such that the action of $S^3_L(p)$ on a light-front spin state gives the rest-frame spin projection along the $z$-direction. As we have shown in Section 3, in fact $\ev{S^3_L(p)}=\ev{S^3}_L(p)=s^3$, and the $z$-component of spin is conserved under Lorentz transformations.

\subsection{ LF conventions}
Light-front coordinates are defined by the light-front time
\begin{align}
x^+= x^0 + x^3
\end{align}
and the corresponding longitudinal spacelike coordinate
\begin{align}
x^-= x^0 - x^3.  
\end{align}
The transverse components $x^1$ and $x^2$ are unchanged  in the light-front coordinates, often denoted by $x^\perp$. Note that by putting $c$ back into the expressions, in the non-relativistic regime when $c \rightarrow \infty$, the light-front time $x^+ = x^0 + \frac{x^3}{c} \rightarrow x^0$ is reduced to the ordinary time. 

The metric tensor is 
\begin{align}
\tensor{g}{^\mu^\nu} = \kbordermatrix{
    & + & - & 1 & 2 \\
     + & 0       & 2	 & 	0  & 0 \\
   - & 2&0 & 0&0 \\
     1& 0       & 0 & -1 & 0\\
      2&  0    & 0 & 0 & -1
} ,  &&\;\;  \tensor{g}{_\mu_\nu} = \kbordermatrix{
    & + & - & 1 & 2 \\[2pt]
     + & 0       & \frac{1}{2}	 & 	0  & 0 \\[2pt]
   - &  \frac{1}{2}	&0 & 0&0 \\
     1& 0       & 0 & -1 & 0\\
      2&  0    & 0 & 0 & -1
}
\end{align}

For any $4-$vecor $v^\mu$, in the light-front coordinates 
\begin{align}
v^\mu= \kbordermatrix{&+&-&1&2\\ &v^+ &v^-&v^1&v^2} ,  &&\;\;v_\mu= \kbordermatrix{&+&-&1&2\\ &\frac{v^-}{2} &\frac{v^+}{2}&-v^1&-v^2}. 
\end{align}

Lorentz invariant scalar product is
\begin{align}
p \cdot x= \frac{p^- }{2} x^+ + \frac{p^+}{2}  x^- - p^1x^1 - p^2x^2.
\end{align} 
$ p^-$ is defined as the light-front Hamiltonian as it is multiplied by $x^+$ the light-front time, $ p^+$ is called the light-front longitudinal momentum for similar reason, and $p^\perp$ is the light-front transverse momentum vector.  For particles on their mass shell, 
\beq
p^-= \frac{{(p^\perp)}^2+ m^2}{p^+}.  
\eeq

Lorentz invariant integral measure is
\begin{align}
\int \frac{\dd^2 p^\perp}{{(2\pi)}^3}   \frac{\dd p^+}{2p^+} =
\int \frac{\dd^4 p}{{(2\pi)}^4} \;\;\theta(p^+)\; (2\pi)\delta^2(p^2-m^2)
\end{align}

\Poincare generators can be classified into the kinematical subgroup which leaves
the quantization plane at one instant of  time invariant, and the dynamical subgroup which involves evolution in time and depends on interactions. In the instant form, the $6$ kinematical generators which leave $x^0=0$ invariant are the translation generators $P^i$ and the rotation generators $J^i$.  In the front form,  there are 
7 kinematical generators under which $x^+=0$:  
\begin{align}
&\text{translation:} &&P^+, P^\perp, \; &&\perp= 1,2  \\
&\text{rotation in the $x-y$ plane:}&&M^{12}\\
&\text{longitudinal boost:} &&M^{+-}= -2M^{03} \\
&\text{transverse boost:} &&M^{+\perp}= M^{0\perp}+ M^{3\perp}, \; &&\perp= 1,2 
\end{align}
The remaining  $3$ dynamical generators   in the front form are
\begin{align}
&\text{Hamiltonian:} &&P^-\\
&\text{transverse rotation:} &&M^{-\perp}= M^{0\perp}- M^{3\perp}, \; &&\perp= 1,2 
\end{align}

Note that as we have shown in Section 3, any transformation generated by the kinematical generators in the front form preserves $j^3$, the angular momentum in the $z$-direction . This is in contrast to the instant form, where  $j^3$ generally even changes under rotations generated by the kinematical operators $J^i$.

\section{Spin Representation in the Front form}
In this section, we derive the front form representations for  spin-$1$ and spin-$\frac{1}{2}$ particles.

Due to Wigner's theorem,  under a Lorentz transformation $\Lambda$, a generic  field $\phi_s(x)$ with spin $s$ transforms in the unitary irreducible representation $\Lambda_s$:
\begin{align}
\phi_s(x) \rightarrow \phi_s'(x)&= \Lambda^{-1}\,\phi_s(x)\,\Lambda \label{99} \\
&=\Lambda_s \; \phi_s(\Lambda^{-1}x) \label{99_1}.
\end{align}

The light-front mode expansion for $\phi_s(x)$ reads: 
\beq
\phi_s(x)= \sum_{\lambda}\int \frac{\dd^2 p^\perp}{{(2\pi)}^{3}} \frac{\dd p^+}{\sqrt{2p^+}} \; \left[\,\mathcal{R}_{s, \lambda}( p) b_\lambda (p) e^{ -i p\cdot x}+ \mathcal{\bar R}_{s,\lambda}( p) d^\dagger_\lambda (p) e^{ i p\cdot x} \, \right].
\eeq
$\lambda$ is as defined in Eq.(\ref{so3}) and (\ref{so2}) for massive and massless particles respectively. $d^\dagger_\lambda (p)$ is the creation operator for particles   
and $ b_\lambda (p)$  is the annihilation operator for anti-particles.  $\mathcal{R}_{s, \lambda}( p)$ and $\mathcal{\bar R}_{s,\lambda}$ represent the spin dependence for the field.  For example, $\mathcal R_0( p)=1$ for scalar particles, $\mathcal R_{1,\pm}(p)= \varepsilon^\mu_{\pm }(p)$ for massless photons,  $\mathcal R_{{\frac{1}{2}}, \pm}(p)= u_{\pm}(p)$ for Dirac spinors, etc.  

Under Lorentz transformation in Eq.(\ref{99_1}),
\beq
\phi'_s(x)= \sum_{\lambda}\int \frac{\dd^2 p^\perp}{{(2\pi)}^{3}} \frac{\dd p^+}{\sqrt{2p^+}} \;  \left[\,\Lambda_s \mathcal{R}_{s, \lambda}( p) b_\lambda (p) e^{ -i (\Lambda p)\cdot x}+ \Lambda_s \mathcal{\bar R}_{s,\lambda}( p) d^\dagger_\lambda (p) e^{ i (\Lambda p)\cdot x} \, \right],\label{102}
\eeq
where we have used $p\cdot (\Lambda^{-1} x) =(\Lambda p)\cdot x $.

For a state $\ket{p; \lambda}= {(2\pi)}^3 \sqrt{2p^+} b^\dagger_\lambda (p) \ket{0}$ in the second quantized form,  it has the corresponding first-quantized wavefunction given by
\beq
\bra{0} \phi_s(x) \ket{p; \lambda}= \mathcal{R}_{s, \lambda}( p)  e^{ -i  p\cdot x}.
\eeq 
Then, under a Lorentz transformation $\ket{p} \rightarrow \Lambda\ket{ p}=\ket{\Lambda p}$, 
\begin{align}
\bra{0} \phi_s(x) \ket{\Lambda p; \lambda}= \mathcal{R}_{s, \lambda}(\Lambda p)  e^{ -i (\Lambda p)\cdot x}. \label{104}
\end{align}
On the other hand, 
\begin{align}
\bra{0} \phi_s(x) \Lambda \ket{ p; \lambda}&= \bra{0} \Lambda^{-1} \phi_s(x) \Lambda \ket{ p; \lambda}\\
&=\bra{0} \phi'_s(x) \ket{ p; \lambda}\\
&= \Lambda_s \mathcal{R}_{s, \lambda}( p)  e^{ -i (\Lambda p)\cdot x}.\label{107}
\end{align}
The first equality is true by assuming the vacuum is Lorentz invariant and $\Lambda\ket{0}=\ket{0}$. The second and third line are obtained using Eq.(\ref{99}) and (\ref{102}).
Equating Eq. (\ref{104}) and (\ref{107}), we derive the Lorentz transformation rule for the spin representation: 
\begin{align}
\mathcal{R}_{s, \lambda}(\Lambda p) =\Lambda_s \mathcal{R}_{s, \lambda}( p).
\end{align}

In the following, we shall start with the spin representations in the standard reference frame,  
and apply this rule to obtain light-front spin representations for any momentum $p$.

\subsection{Spin-$1$ with $m\neq0$}

A massive spin-$1$  field has the mode expansion
\beq
B^\mu(x)= \sum_{\lambda=-1,0,1}\int \frac{\dd^2 p^\perp}{{(2\pi)}^{3}} \frac{\dd p^+}{\sqrt{2p^+}} \; \left[\,\varepsilon^\mu_\lambda( p) a_\lambda (p) e^{ -i p\cdot x}+ \varepsilon^{\ast\mu}_\lambda( p) a^\dagger_\lambda (p) e^{ i p\cdot x} \, \right].
\eeq

The standard reference frame for massive particles is the particle's rest frame defined in Eq.(\ref{so3}) with $\mathring p^\mu= \kbordermatrix{&+&-&1&2\\ &m &m&0&0}$. The  polarization vectors in this frame correspond to the eigenvectors of the little group $SO(3)$ with $\lambda=\pm 1,0$:
\begin{align}
\varepsilon^\mu_+(\mathring p)=  \kbordermatrix{& &\\ + & 0 \\[1pt] - &0  \\[1pt] 1 & \frac{-1}{\sqrt 2}  \\[5pt]2& \frac{-i}{\sqrt 2}} &&
\varepsilon^\mu_-(\mathring p)=  \kbordermatrix{& &\\[1pt] + & 0\\ - &0 \\[1pt] 1 & \frac{1}{\sqrt 2} \\[4pt] 2& \frac{-i}{\sqrt 2}} &&
\varepsilon^\mu_0(\mathring p)=  \kbordermatrix{& &\\[1pt] + & 1\\ - &1 \\[1pt] 1 &0 \\[4pt] 2& 0} 
\end{align}
in the light-front coordinates.
Applying the vector representation of the light-front boost $\tensor{\left(\Lambda_L\right)}{^\mu_\nu}(\mathring p \rightarrow p)$ in Eq.$(\ref{vecl})$, we find 
\begin{align} 
\varepsilon^\mu_+(p)=  \kbordermatrix{& &\\ + & 0\\[3pt] - &\frac{- \sqrt{2}\;p^\mathrm R }{p^+}  \\[5pt]1 & \frac{-1}{\sqrt 2} \\[5pt] 2& \frac{-i}{\sqrt 2}} &&
\varepsilon^\mu_-( p)=  \kbordermatrix{& &\\ + & 0\\[3pt] - &\frac{\sqrt{2}\;p^\mathrm L }{p^+}  \\[5pt] 1 & \frac{1}{\sqrt 2} \\[5pt] 2& \frac{-i}{\sqrt 2}}&&
\varepsilon^\mu_0( p)=  \kbordermatrix{& &\\ + & \frac{p^+}{m}\\[3pt] - &\frac{|p^\perp|^2 + m^2}{m} \\[5pt] 1 & \frac{p^1}{m}\\[5pt] 2& \frac{p^2}{m}} \label{111},
\end{align}
where  $p^\mathrm R \equiv p^1+ i p^2$ and $p^\mathrm L \equiv  p^1- i p^2$. $\varepsilon^\mu_+(p)$ and $\varepsilon^\mu_-(p)$ are sometimes referred to as the right-handed and left-handed circular polarization respectively.

\subsection{Spin-$1$ with $m=0$}
A massless spin-$1$  field has the mode expansion
\beq
A^\mu(x)= \sum_{\lambda=\pm}\int \frac{\dd^2 p^\perp}{{(2\pi)}^{3}} \frac{\dd p^+}{\sqrt{2p^+}} \; \left[\,\varepsilon^\mu_\lambda( p) a_\lambda (p) e^{ -i p\cdot x}+ \varepsilon^{\ast\mu}_\lambda( p) a^\dagger_\lambda (p) e^{ i p\cdot x} \, \right].
\eeq

The standard reference frame for massless particles is defined in Eq.(\ref{so2}), in which a particle moves along the $z$-direction with $\bar p^\mu= \kbordermatrix{&+&-&1&2\\ &2\bar p&0&0&0}$. The polarization vectors in this frame are the eigenvectors of the little group  $SO(2)$  with $\lambda=\pm 1$:
\begin{align}
\varepsilon^\mu_+(\bar p)=  \kbordermatrix{& &\\ + & 0 \\[1pt] - &0  \\[1pt] 1 & \frac{-1}{\sqrt 2}  \\[5pt]2& \frac{-i}{\sqrt 2}} &&
\varepsilon^\mu_-(\bar p)=  \kbordermatrix{& &\\[1pt] + & 0\\ - &0 \\[1pt] 1 & \frac{1}{\sqrt 2} \\[4pt] 2& \frac{-i}{\sqrt 2}} 
\end{align}
in the light-front coordinates.
Applying the vector representation of the light-front boost $\tensor{\left(\Lambda_L\right)}{^\mu_\nu}(\bar{p} \rightarrow p)$ in Eq.$(\ref{vec_l})$, we find

\begin{align} 
\varepsilon^\mu_+(p)=  \kbordermatrix{& &\\ + & 0\\[3pt] - &\frac{- \sqrt{2}\;p^\mathrm R }{p^+}  \\[5pt]1 & \frac{-1}{\sqrt 2} \\[5pt] 2& \frac{-i}{\sqrt 2}} &&
\varepsilon^\mu_-( p)=  \kbordermatrix{& &\\ + & 0\\[3pt] - &\frac{\sqrt{2}\;p^\mathrm L }{p^+}  \\[5pt] 1 & \frac{1}{\sqrt 2} \\[5pt] 2& \frac{-i}{\sqrt 2}}
\label{124}
\end{align}
The above expressions are consistent with the transverse polarizations $\epsilon^\mu_{\pm}(p)$ obtained for massive particles in Eq.(\ref{111}). 

Note that the $+$ component of the polarization vectors vanishes for all momentum $p$. Therefore, the light-front gauge condition $A^+=0$ is preserved under the light-front Lorentz transformation $\Lambda_L$. This is in contrast to other choices of Lorentz transformations, under which a gauge condition is generally not preserved.

\subsection{Spin-$\frac{1}{2}$ Dirac spinors}
A Dirac spin-$\frac{1}{2}$ fermion has the mode expension
\beq
\psi(x)= \sum_{\lambda=\pm \frac{1}{2}}\int \frac{\dd^2 p^\perp}{{(2\pi)}^{3}} \frac{\dd p^+}{\sqrt{2p^+}} \; \left[\, u_\lambda ( p) b_\lambda (p) e^{ -i p\cdot x}+ v_\lambda ( p) d^\dagger_\lambda (p) e^{ i p\cdot x} \, \right].
\eeq

Since both massive and massless Dirac fermions have two polarizations, the spin representations must be consistent in both cases. We shall work with the massive case and obtain the expression for massless Dirac spinors by taking $m\rightarrow 0$. 

In a massive Dirac fermion's rest frame, solutions to the Dirac equation correspond to eigenvectors of the $SO(3)$ rotation group with $\lambda=\frac{1}{2}, -\frac{1}{2}$, labeled by $\uparrow$ and $\downarrow$ respectively:
\begin{align}
u_{\uparrow}(\mathring p)= \sqrt{2m}\begin{bmatrix} 1 \\ 0 \\ 0\\ 0 \end{bmatrix} &&  
u_{\downarrow}(\mathring p)= \sqrt{2m}\begin{bmatrix} 0 \\ 1 \\ 0\\ 0 \end{bmatrix} &&
v_{\uparrow}(\mathring p)= \sqrt{2m}\begin{bmatrix} 0 \\ 0 \\ 0\\ -1 \end{bmatrix} &&
v_{\downarrow}(\mathring p)= \sqrt{2m}\begin{bmatrix} 0 \\ 0 \\ 1\\ 0 \end{bmatrix}.
\end{align}
The spinors are defined in the Dirac representation, where
\begin{align}
\gamma^0= \begin{bmatrix}
    \mathbbm{1}      &     0 \\
    0        &  - \mathbbm{1}
\end{bmatrix} && 
\gamma^i= \begin{bmatrix}
    0     &     \sigma^i \\
    -\sigma^i        &  0
\end{bmatrix} &&
 \gamma^5= \begin{bmatrix}
    0     &     \mathbbm{1}\\
    \mathbbm{1}        &  0
\end{bmatrix} 
\end{align}

The Lorentz transformations for spin-$\frac{1}{2}$ are generated by \\$
S^{\mu\nu}= \frac{i}{4} [\gamma^\mu, \gamma^\nu]=\frac{1}{2}\sigma^{\mu\nu}.$ In the Dirac representation, the light-front boost generators are
\begin{align}
S^{+-}= -i \begin{bmatrix}
    0   &     \sigma^3 \\
     \sigma^3         &  0
\end{bmatrix} && 
S^{+1}= \frac{1}{2}\begin{bmatrix}
    \sigma^2    &     i\sigma^1 \\
    i\sigma^1        &  \sigma^2
\end{bmatrix} &&
 S^{+2}= \frac{1}{2} \begin{bmatrix}
    -\sigma^1    &     i\sigma^2\\
     i\sigma^2        &  -\sigma^1
\end{bmatrix} 
\end{align}

Applying to the light-front boost in  Eq.$(\ref{lal})$, we obtain the spin-$\frac{1}{2}$ representation 
\begin{align}
\Lambda_{L,\frac{1}{2}}(\mathring{p} \rightarrow p) =\sqrt{\frac{1}{4mp^+}}\begin{bmatrix}
      p^+ + m   & -p^L & p^+ - m  & p^L    \\\\
       p^R & p^+ + m  & p^R & -p^++ m   \\\\
       p^+ - m   & p^L & p^+ + m  & -p^L    \\\\
       p^R & -p^+ + m  & p^R & p^++ m 
\end{bmatrix},
\end{align}
where $ p^R= p^1 + i p^2 $ and $p^L= p^1 - i p^2.$

The light-front spinors at any momentum are thus given by \cite{Lepage:1980fj}
\begin{align}
u_{\uparrow}(p)= \sqrt{\frac{1}{2p^+}}\begin{bmatrix} p^+ + m \\[5pt] p^R \\[5pt] p^+ - m\\[5pt] p^R \end{bmatrix} &&  
u_{\downarrow}(p)= \sqrt{\frac{1}{2p^+}}\begin{bmatrix} -p^L \\[5pt] p^+ + m \\[5pt]  p^L \\[5pt]-p^++ m\end{bmatrix} \label{176}
\end{align}
\begin{align}
v_{\uparrow}(p)= \sqrt{\frac{1}{2p^+}}\begin{bmatrix} -p^L \\[5pt] p^+ - m \\[5pt]  p^L \\[5pt] -p^+ - m\end{bmatrix} &&  
v_{\downarrow}(p)= \sqrt{\frac{1}{2p^+}}\begin{bmatrix} p^+ - m \\[5pt] p^R \\[5pt] p^+ + m\\[5pt]p^R \end{bmatrix}. \label{177}
\end{align}
It is interesting that the light-front spinors are exactly what have been used in the field of scattering amplitudes \cite{Dixon:1996wi}.

\section{Pauli-Lubanski pseudovectors and Spin}
In this section, we derive the relations between the Pauli-Lubanski pseudovector and various relativistic expressions of spin.

\subsection{Covariant spin sector $s^\mu(p)$}

In nonrelativistic physics, for spin-$s$ particles,  the three components of a spin vector $s^i$  is defined as the expectation value of the the spin generators $S^i$:
\begin{align}
 s^i =
 \begin{bmatrix}
s^1\\
s^2\\
s^3
\end{bmatrix}= \frac{1}{s}\,\bra{ s^i}  \begin{bmatrix}
S^1\\
S^2\\
S^3
\end{bmatrix}  \ket{ s^i}.
\end{align}
In the main text, we have chosen to label spin in the $z$-direction and set $s^1=s^2=0$. But labeling spin with the other two axes will not affect the discussion as long as all the operators are redefined accordingly.

The definition of a spin vector can be extended to relativistic systems  via the Pauli-Lubanski pseudovector 
\beq
W^\mu= -\frac{1}{2} \varepsilon^{\mu\nu\alpha\beta} { P}_\nu { M}_{\alpha\beta}.  
\eeq

For a particle moving with momentum $p$, a relativistic spin vector ${s}^\mu(p)$ is defined as 
\begin{align}
 {s}^\mu(p)=\frac{1}{s }\,\frac{\bra{p; s^i} W^\mu \ket{p; s^i}}{\braket{p; s^i|p; s^i}}.
\end{align}
It is easy to check that the $s^\mu(p)$ is consistent with the nonrelativistic spin vector $s^i$ by going to a massive particle's rest frame. In the rest frame, $W^\mu$ is proportional to the $SO(3)$ rotation generators as in Eq.(\ref{1}), and 
\begin{align}
 {s}^\mu(\mathring p)= \kbordermatrix{& &\\ 0 & 0 \\ i &\frac{m}{s} s^i}.  
\end{align}

Furthermore, $s^\mu(p)$ transforms covariantly  under Lorentz transformations:  Under a Lorentz transformation $\Lambda$, $\Lambda^{-1}\,P^\mu \,\Lambda = \tensor{\left(\Lambda\right)}{^\mu_\nu} P^\nu$ and
$\ket{p} \rightarrow \Lambda\ket{ p}=\ket{\Lambda p}$. Since  $W^\mu$ satisfies $W \cdot P=0$,  $W^\mu$ must also transform as a $4-$vector:
\begin{align}
\Lambda^{-1}\,W^\mu \,\Lambda &= \tensor{\left(\Lambda\right)}{^\mu_\nu} W^\nu. \label{137}
\end{align}
Then, the spin vector 
\begin{align}
 {s}^\mu(p) \rightarrow  {s}^\mu(\Lambda p) &=\frac{1}{s } \frac{\bra{\Lambda p; s^i} W^\mu \ket{\Lambda p; s^i}}{\braket{\Lambda p; s^i|\Lambda p; s^i}}\\
 &=  \frac{1}{s } \frac{\bra{ p; s^i}\tensor{\left(\Lambda\right)}{^\mu_\nu}  W^\nu  \ket{ p; s^i}}{\braket{\Lambda p; s^i|\Lambda p; s^i}}\\
 &=\tensor{\Lambda}{^\mu_\nu}\, s^\nu(p).
\end{align}
The last line is true because the state normalization is Lorentz invariant and ${\braket{\Lambda p; s^i|\Lambda p; s^i}}={\braket{ p; s^i| p; s^i}}= 2p^+ {(2\pi)}^3 \delta^3(0)$.

We can thus obtain the spin vector in any frame by transforming it as a vector from the rest frame. In the instant form, the covariant spin vector is naturally defined with the canonical choice of Lorentz transformation in Eq.(\ref{vecc}) as it is smoothly connected to identity in the nonrelativistic regime, and
\beq
s^\mu_c(p)=
\kbordermatrix{& &\\ 0 & p^js^j  \\[5pt] i &m s^i + \frac{ p^j s^j\, }{p^0+m}p^i } \frac{1}{s}
\eeq

In the front form, the spin vector is defined with the light-front boost in Eq.(\ref{vecl}), and 
\beq
s^\mu_L(p)=
\kbordermatrix{& &\\ + & p^+ s^3 \\[3pt] - & \frac{\abs{p^\perp}^2- m^2}{p^+}s^3 + \frac{2m p^\perp s^\perp}{p^+}
\\[6pt] \perp & m s^\perp + s^3 p^\perp } \frac{1}{s}.
\eeq

 \subsection{Light-front spin operator $S^i_L(p)$ }
In addition to the light-front spin operator $S^3_L(p)$ in the $z$-direction  in Eq.( \ref{s3p}), one can also define transverse light-front spin operators $S^\perp_L(p)$ analogously such that they give $s^\perp$, the angular momentum projection along the transverse direction defined in the standard reference frame. We then have 
\begin{align}
&{S^i_L}(p) \ket{p; j^i}_L = j^i  \ket{p; j^i}_L \label{174}\\[5pt] 
\text{with\;\;} &{S^i_L}(p) = \Lambda_L(\mathring{p} \rightarrow p) \; J^i  \; {\Lambda^{-1}_L}(\mathring{p} \rightarrow p), \;\; i=1,2,3.
\end{align}
For an elementary particle, $j^i= s^i$, and for a composite system with $n$ constituents, $\displaystyle j^i= \sum_{a=1}^n \;  s^i_a+ \sum_{a=1}^{n-1} \;l^i_a$. 

Let us relate the light-front spin operators  to the Pauli-Lubanski vector. Under the light-front Lorentz transformation defined in Eq.(\ref{vecl}), 
\begin{align}
W^\mu= \kbordermatrix{& &\\ + & W^+  \\ - & W^- \\ \perp &  W^\perp } 
&= \tensor{\left(\Lambda_L\right)}{^\mu_\nu} \Lambda_L \,W^\nu \,\Lambda^{-1}_L \\
 &= \tensor{\left(\Lambda_L\right)}{^\mu_\nu} \Lambda_L \,  \begin{bmatrix} mJ^3 \\ -mJ^3\\ mJ^\perp \end{bmatrix}\Lambda^{-1}_L \\[4pt]
 &= \tensor{\left(\Lambda_L\right)}{^\mu_\nu} \begin{bmatrix} mS^3_L(p) \\[3pt] -mS^3_L(p) \\[3pt] mS^\perp_L(p) \end{bmatrix}\\[6pt]
 &= \begin{bmatrix} P^+ S^3_L(p) \\[3pt] \frac{\abs{P^\perp}^2- m^2}{P^+}S^3_L(p) + \frac{2m P^\perp S^\perp_L(p)}{P^+} \\[6pt]  mS^\perp_L(P)+ S^3_L(p) P^\perp \end{bmatrix}.
\end{align}
We then recover the familiar expression\cite{Leutwyler:1977vy}:
\begin{align}
S^3_L(p)&= \frac{W^+}{P^+} \;\; &&
m S^\perp_L(p)= W^\perp - S^3_L(p)P^\perp.
\end{align}

In a similar way, one can also check that the Wick helicity operator $S^3_h(p)$ defined in Eq.(\ref{64}) is related to the Pauli-Lubanski vector by 
\beq
S^3_h(p)= \frac{W^0}{\abs{\mathbf p}},
\eeq
and the helicity operator thus always only depends on the $3$-momentum $\mathbf p$.

\end{appendices}
\cleardoublepage

\bibliographystyle{utphys}
\bibliography{j_z}

\end{document}